\documentclass[aps,prb,preprint,groupedaddress,showpacs,showkeys]{revtex4}

\usepackage{graphicx}

\begin{document}


\title{SQUID magnetometry of superconducting samples: The case of RuSr$_{2}$GdCu$_{2}$O$_{8}$}


\author{Thomas P. Papageorgiou}
\email[Corresponding author: ]{Thomas.Papageo@uni-bayreuth.de}
\author{Ludwig Bauernfeind}
\author{Hans F. Braun}
\affiliation{Physikalisches Institut, Universit\"at Bayreuth, D-95440 Bayreuth, Germany}


\date{\today}

\begin{abstract}

Inhomogeneities of the field in the superconducting magnet of a 
superconducting quantum intereference device (SQUID) magnetometer 
can create serious artefacts in the d.c. magnetization measurements of superconducting samples. 
We discuss the problem focusing on the procedure of calculating the magnetic moment of a 
sample from the measured SQUID output signal. The (weakly) ferromagnetic superconductor 
RuSr$_{2}$GdCu$_{2}$O$_{8}$ has been reported sensitive to inhomogeneities of the SQUID's 
superconducting magnet, which create artefacts in its measured magnetization. 
Indeed, we show that, because of the small values of the magnetic hysteresis width, 
field changes less than 100~$\mu$T over the scaning 
length inside the magnet are enough to create spurious signals in the measured magnetic 
moment of this compound.  

\end{abstract}

\pacs{74.72.-h, 74.25.Ha, 74.62.Yb}
\keywords{superconductivity, SQUID magnetometry , RuSr$_{2}$GdCu$_{2}$O$_{8}$}

\maketitle

\section{Introduction}

A common method to calculate the magnetic moment of a sample necessitates its movement through 
a pick-up coil system inside the magnet of a superconducting quantum intereference device (SQUID) magnetometer. The obtained 
response, i.e. the pick-up coil signal as a function of the sample's position on the 
axis of the pick-up coil system, is fitted using the theoretical response of a magnetic 
point dipole of moment \textit{m}. The fitting parameter \textit{m}
is compared with the system calibration and the sign and value of the sample's magnetic moment 
is calculated. 
  
The above method for the determination of the magnetic moment can sometimes create serious 
problems in the measurements of superconducting samples, when these are done in a non-homogeneous field. 
This is because most of the analysis methods of the SQUID's response assume that the magnetic 
moment of the sample does not change during the measurement. A superconducting sample though, will 
follow a minor hysteresis loop during the measurement, if this is done in a non-homogeneous field. 
Thus, the value of the sample's magnetic moment calculated by the magnetometer's software will not 
represent the actual moment of the sample at the temperature of the measurement. 

Such problems, which may create spurious features in the d.c. magnetization of superconducting materials, 
have been discussed by several authors in the past. For example, Blunt \textit{et al.} \cite{Blunt} 
investigating the superconducting properties of (TlV$_{0.5}$)Sr$_{2}$(Ca$_{0.8}$Y$_{0.2}$)Cu$_{2}$O$_{y}$ suggest, 
that the paramagnetic moments 
reported by Lee \textit{et al.} \cite{Lee} are artefacts arising from the movement of the sample in a 
non-homogeneous field during the measurements. Schilling \textit{et al.} \cite{Schilling} show, that the 
\textit{M}(\textit{T},\textit{H}) measurement of YBCO single crystals at temperatures below the 
irreversibility temperature is seriously affected by a spatial inhomogeneity of the applied field and 
Braunisch \textit{et al.} \cite{Braunisch}, who discovered the Paramagnetic Meissner Effect in 
Bi High Temperature Superconductors, point out the necessity, that the samples do not move 
during the measurement for reliable results to be produced. 

In this paper, we present the problem of SQUID magnetometry for superconducting samples in more 
detail. We show how the measured signals can be deformed when a measurement is done in 
a non-homogeneous field and point out the errors that can occur when the magnetic moment of 
a superconducting sample is extracted from these deformed signals. As an example we discuss the 
case of RuSr$_{2}$GdCu$_{2}$O$_{8}$ (Ru-1212). It has recently been shown \cite{Papageorgiou}, 
that this weakly ferromagnetic superconductor is sensitive to field inhomogeneities, which can create 
problems to the measurements of its magnetic moment. We use the Bean model to show, that field 
changes less than 100~$\mu$T over the scanning length in the SQUID magnetometer can be enough to create artefacts 
in the measured magnetization of Ru-1212. 

\section{SQUID magnetometry of superconducting samples}

\subsection{How does a SQUID magnetometer measure ?}

For many of the commercially available magnetometers the measurement requires the motion of 
the sample through a pick-up coil system. The coils are wound in a second derivative configuration, 
where the two outer detection loops, located at a distance \textit{A} from the center of the 
magnetometer's magnet, are wound oppositely to the two central loops located at the center of the 
magnet. During the measurement at a certain temperature, the movement of the sample through 
the pick-up coils induces currents in the detection loops, which, through an inductance \textit{L}, 
create magnetic flux in the SQUID circuit, resulting in an output voltage \textit{V}, which 
depends on the position of the sample \textit{z}. 

\begin{figure}
\includegraphics[clip=true,width=75mm]{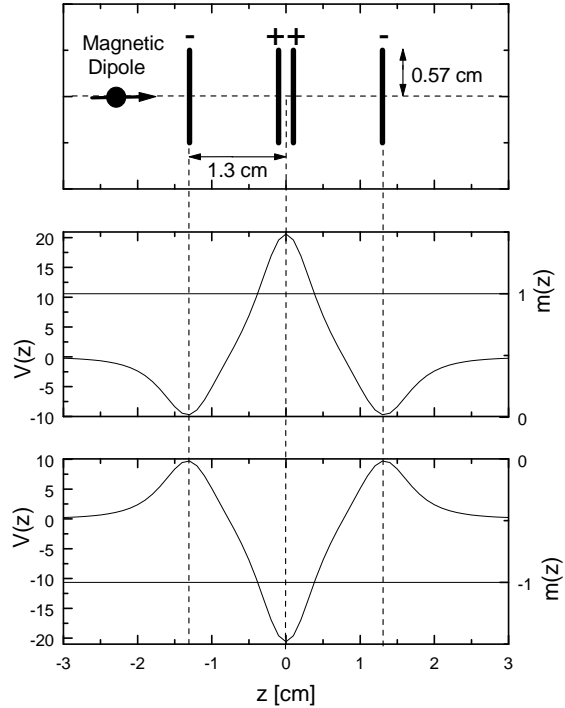}
\caption{Schematic presentation of the pick-up coil system in a SQUID magnetometer with 
the two central loops (+) wound oppositely to the two outer loops (-) (top). The theoretical 
response of this system to the movent of a point dipole of constant positive (middle) and 
negative moment (bottom) is also given.}\label{fig:1}
\end{figure}

It is rather trivial to show, that the magnetic flux $\Phi$ through a loop with radius \textit{R} created by 
a point dipole of magnetic moment \textit{m} located at a distance \textit{D} from the loop on 
the symmetry axis is:  

\begin{equation}
\Phi = \frac{2 \pi m R^{2}}{(D^{2}+R^{2})^{3/2}} \label{eq:dresp}
\end{equation} 
\linebreak
Thus, the response of a pick up coil system like the one described above to the movement of a 
point dipole on its axis \textit{z} will be proportional to: 

\begin{equation}
V(z) = 2 \pi m R^{2}\left(\frac{2}{[z^{2}+R^{2}]^{3/2}}-\frac{1}{[(z+A)^{2}+R^{2}]^{3/2}}-\frac{1}{[(z-A)^{2}+R^{2}]^{3/2}}\right) \label{eq:SQUID}
\end{equation}
\linebreak
where \textit{z} is the position of the sample. In Eq.~\ref{eq:SQUID} the two central loops of the 
pick-up coil system are assumed to be at \textit{z}=0 while the two oppositely wound outer loops 
at \textit{z}=\textit{A} and \textit{z}=-\textit{A} respectively. In figure~\ref{fig:1} we show 
the pick-up coil voltage for a point dipole with positive (middle) and negative (bottom) 
moment constant over the scaning length. Note that in this case the signal is very well defined with 
extrema at the position of the central and outer loops.

During the measurement of a sample, signals similar to those presented in figure~\ref{fig:1} 
are measured. In this case, the contributions to the measured signal can have more than one source 
(sample, sample holder, rod where the sample holder is mounted e.t.c.). Nevertheless, Eq.~\ref{eq:SQUID} 
is used to fit the contribution from the sample and calculate its magnetic moment \textit{m}. This is 
because nearly all analysis methods of the measured \textit{V}(\textit{z}) signal make two significant 
assumptions for the magnetic moment of the sample (i) that it approximates a magnetic dipole moment and 
(ii) that the sign and value of this moment do not change during the measurement.

Practically, the above assumptions mean, that the magnetometer's software will be able to fit properly the
measured signals which are similar to the well defined signals of figure~\ref{fig:1}. If, for any reason, the 
magnetic moment of the sample does not approximate a magnetic dipole moment or if this moment 
is changing over the scaning length, then the measured signal will be distorted compared to that of 
figure~\ref{fig:1}, the produced value of the magnetic moment from the magnetometer's sofware will 
come from the best possible fitting to this distorted signal and it will probably deviate from the 
real value of the sample's magnetic moment in the temperature of the measurement. 

\subsection{Measuring a superconductor}  

\begin{figure}
\includegraphics[clip=true,width=75mm]{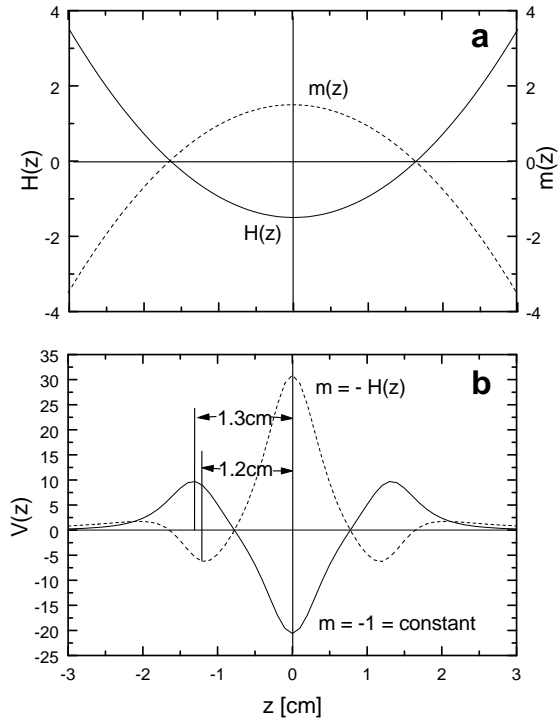}
\caption{(a) The magnetic moment \textit{m}(\textit{z}) of a superconducting sample in the Meissner state 
which is moved over the field profile \textit{H}(\textit{z}). (b) Calculated response (dashed line) of 
the pick-up coil system for the magnetic moment \textit{m}(\textit{z}). The ideal response for a sample of 
constant negative moment is also shown (solid line).}\label{fig:2}
\end{figure}

In order to illustrate the problems that field inhomogeneities can cause to the SQUID magnetitzation 
measurements of a superconductor, we will investigate the case of a sample in the Meissner state 
measured in a field, which over the scanning length in the pick-up coil system has the profile shown 
in figure~\ref{fig:2}a. Such a profile, which changes sign over the scanning length, can represent 
the profile of the remanent field in the superconducting magnet \cite{McElfresh} and measurements in 
set (positive) values of the magnetic field of the order of the remanent field will probably retain a profile 
similar to that of figure~\ref{fig:2}a.  

\begin{figure}
\includegraphics[clip=true,width=75mm]{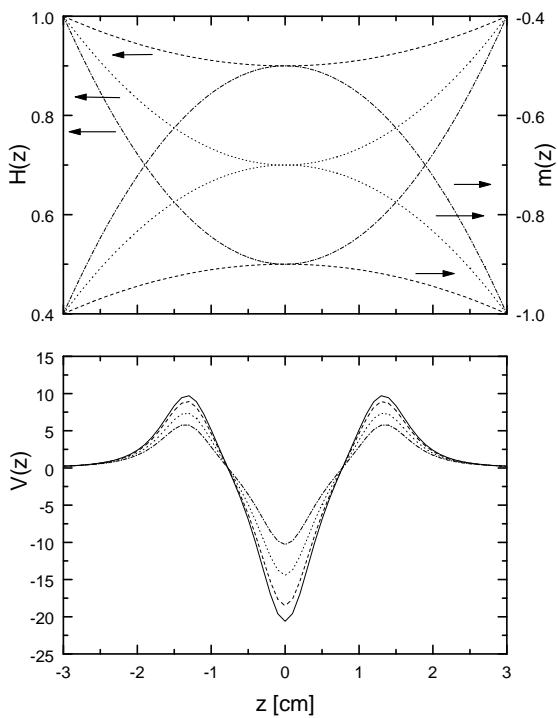}
\caption{(top) magnetic moment \textit{m}(\textit{z}) of a superconductor in the 
Meissner state for field profiles with different level of homogeneity. (bottom) 
Calculated response of the pick-up coil system for the above \textit{m}(\textit{z}). The ideal 
response for \textit{m}=-1=constant is also shown (solid line).}\label{fig:3}
\end{figure}

Such field profiles will cause a change of the magnetic moment \textit{m}(\textit{z}), as shown in 
figure~\ref{fig:2}a, thus, the assumption for constant magnetic moment will be violated. In 
figure~\ref{fig:2}b we show the response of the pick-up coil system to the magnetic 
moment \textit{m}(\textit{z}). For comparison the ideal response for a constant negative moment is 
also shown. The most striking mistake that the magnetometer's software will do by fitting the signal 
due to the non-constant moment is, that it will produce a positive value for the magnetic moment of the 
sample although this is a superconductor in the Meissner state.

Note that if no information about the field profile and the nature of the sample is available, it is 
sometimes very difficult to notice the artefact from the measured signal. The signal of figure~\ref{fig:2}b 
corresponding to \textit{m}(\textit{z}) looks very similar to the ideal signals of figure~\ref{fig:1} and 
would probably not warn the user of the SQUID magnetometer for a possible artefact in the measurement. A 
careful inspection of the curve though, shows that the two minima are not located at the position of the 
outer coils but 0.1 cm closer to \textit{z}=0 (see figure~\ref{fig:2}b). Thus, very careful analysis of the 
measured signals is necessary for safe conclusions to be drawn, since sometimes the artefact is hidden in 
small distortions of the signal (the crossover points are slightly shifted like in figure~\ref{fig:2}b or 
the shape of the scan is not quite right) and as it was already pointed out by Blunt \textit{et al.} \cite{Blunt} 
most users of SQUID magnetometers, and especially the inexperienced ones, will miss this fine detail. 

In the case, where the field profile is not changing sign, the 
fitting of the measured signal will give the correct sign for the magnetic moment of the sample. 
Nevertheless, since, as shown in figure~\ref{fig:3}, the higher the level of the field inhomogeneity, the bigger 
the distortion of the measured signal compared to that corresponding to \textit{m}=constant, the magnetometer's 
software will calculate values of the magnetic moment which, although they will have the correct sign, 
will not represent the actual value of the magnetic moment for the sample.

\section{The case of $\mbox{RuSr}_{2}\mbox{GdCu}_{2}\mbox{O}_{8}$} 

The weakly ferromagnetic superconductor Ru-1212 has recently drawn a lot of attention since it is 
one of the very few high temperature superconductors, where superconductivity arises in a state, in 
which magnetic order is already developed. Nevertheless, there is some sceptisism whether Ru-1212 
is a bulk superconductor, which, among others, arises from the contradicting results on the d.c. 
magnetization of this compound \cite{comment}. 

\begin{figure}
\includegraphics[clip=true,width=75mm]{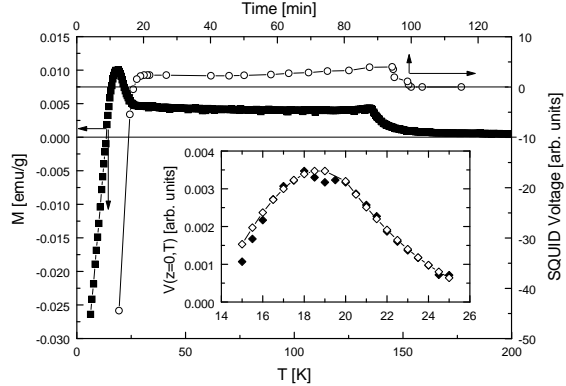}
\caption{Zero field cooled d.c. magnetization measurement of Ru-1212 showing a peak like feature below \textit{T$_{c}$} 
(closed squares). Measuring the output voltage without moving the sample (open circles) no peak like 
feature is observed. The measuring field was 250~$\mu$T. Inset: measured \textit{V}(\textit{z}=0,\textit{T}) signal (closed diamonds). 
The line with the open diamonds represents the fitting of the measured 
\textit{V}(\textit{z}=0,\textit{T}) signal (see text).}\label{fig:4}
\end{figure}

Recently it has been shown \cite{Papageorgiou}, that Ru-1212 is sensitive to field inhomogeneities 
over the scanning length in the SQUID magnetometer, which create artefacts in its measured magnetization.
For example, in figure~\ref{fig:4} we show a zero field cooled (z.f.c.) measurement [the technique for the 
cancelation of the remanent fields of the superconducting magnet as well as the earth field for the achievement of a field value 
close to zero is described in references 5 and 14] on a Ru-1212 sample 
showing a peak-like feature below the superconducting transition temperature \textit{T$_{c}$}=30~K, as 
it was determined by resistivity and a.c. susceptibility measurements \cite{Papageorgiou}. A similar 
feature was also observed in the field cooled (f.c.) curve \cite{Papageorgiou}. The magnetic moment of the 
sample for these measurements was determined by the magnetometer's software from the signal \textit{V}(\textit{z}) 
as described in the previous sections by moving the sample through the pick-up coil system in the superconducting 
magnet of the SQUID magnetometer. In the same figure we 
show the output voltage of the SQUID circuit recorded without moving the sample, which is proportional to any changes 
of the sample's magnetic moment as the temperature is changing. No peak like feature is observed in this curve 
leading us to the conclusion, that this feature is an artefact arising from the movement of the sample in an 
inhomogeneous field during the measurement in the SQUID magnetometer. 
In view of the observed sensitivity of the Ru-1212 measured magnetization in its superconducting state to field 
inhomogeneities, the contradicting results on the measured magnetization in the superconducting state 
of Ru-1212 could be explained as the result of different field profiles in the superconducting magnet of the SQUID 
magnetometer. In the following we will attempt an estimation of the magnitude of the 
field inhomogeneity necessary to create artefacts in the measured d.c. magnetic moment of Ru-1212. 

\begin{figure}
\includegraphics[clip=true,width=75mm]{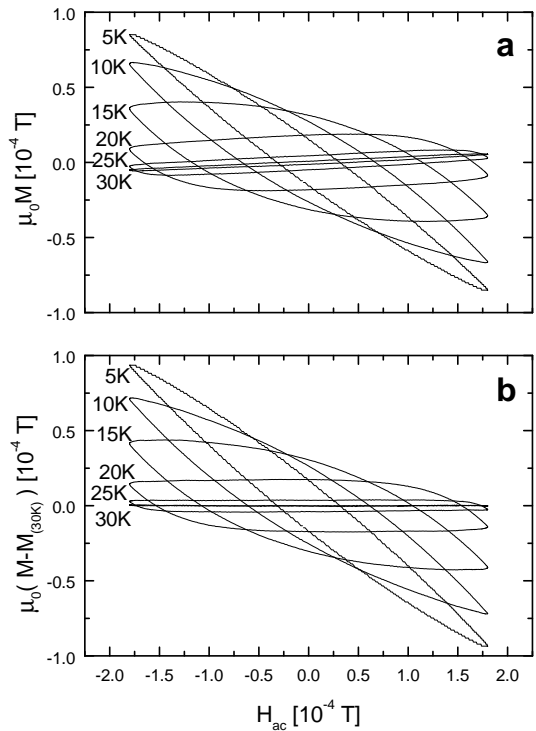}
\caption{(a) Hysteresis loops for our Ru-1212 sample at different temperatures below \textit{T$_{c}$}=30~K 
(b) The same as (a) after the measured curve at 30~K has been subtracted from the data.}\label{fig:5}
\end{figure}

For the study of the effect of magnetic field inhomogeneities on the d.c. magnetization measurements of Ru-1212 
we will need, first of all, a model to describe the field penetration into the sample. Chen \textit{et al.} 
\cite{Chen} have shown, that already from the shape of the \text{M}(\textit{H}) curves, information about the 
applicability of a critical state model can be obtained, since different, characteristic, shapes are expected for 
different models. In figure~\ref{fig:5} we show measured hysteresis loops for a Ru-1212 sample obtained using a 
home made a.c. susceptometer \cite{PhDthesis}. The sample was kept stationary within the pick-up coil system, while the primary 
field is created by a coil made from normal conducting Cu-wire. This way we avoid the problems described above, 
which are related to the movement of the sample in a non-homogeneous field, as well as uncertainties in the determination 
of the field arising from the remanent fields or trapped flux in the superconducting magnet of the SQUID magnetometer. This is 
an important point since the range of fields (less than 200~$\mu$T) scanned are within the range of the remanent field of 
a superconducting magnet. In an attempt to correct for the paramagnetic contribution to 
the signal from the Gd moments we have subtracted the measured curve at 30~K, the shape of which is very close 
to a paramagnetic line (see figure~\ref{fig:5}a), from the other measured curves. The result is shown in 
figure~\ref{fig:5}b. It is obvious from the shape of our curves, that the Bean critical state model is 
appropriate for the description of these magnetization curves \cite{Chen}. We note, that although the Bean 
critical state model is routinely used to describe the properties of many superconducting systems, careful check of 
its applicability is necessary. For example, for Ru-1212 a transition from a Bean model-like behaviour 
to a Kim model-like behaviour has been observed for field changes above $\sim$ 8~Gauss \cite{Bauernfeind}. Nevertheless, 
as shown above, for fields in the range shown in figures~\ref{fig:4}-~\ref{fig:5} the Bean critical state model can be used.

Libbrecht \textit{et al.} \cite{Libbrecht} have applied the Bean model in order to describe the influence of field 
inhomogeneities on the measured magnetization of YBCO films. They model the temperature dependence of the penetration field 
by a power law of the type:

\begin{equation}
H^{*}=H_{0}^{*}\left(1-\frac{T}{T_{ir}}\right)^{n} \label{eq:3}
\end{equation} 
where $H^{*}$ is the lowest applied field for full flux penetration and $T_{ir}$ is the irreversibility temperature, and 
present equations to fit their measured \textit{V}(\textit{z},\textit{T}) signals, assuming a field profile similar to 
the one shown in figure~\ref{fig:6}, which is described by the formula: 

\begin{figure}
\includegraphics[clip=true,width=75mm]{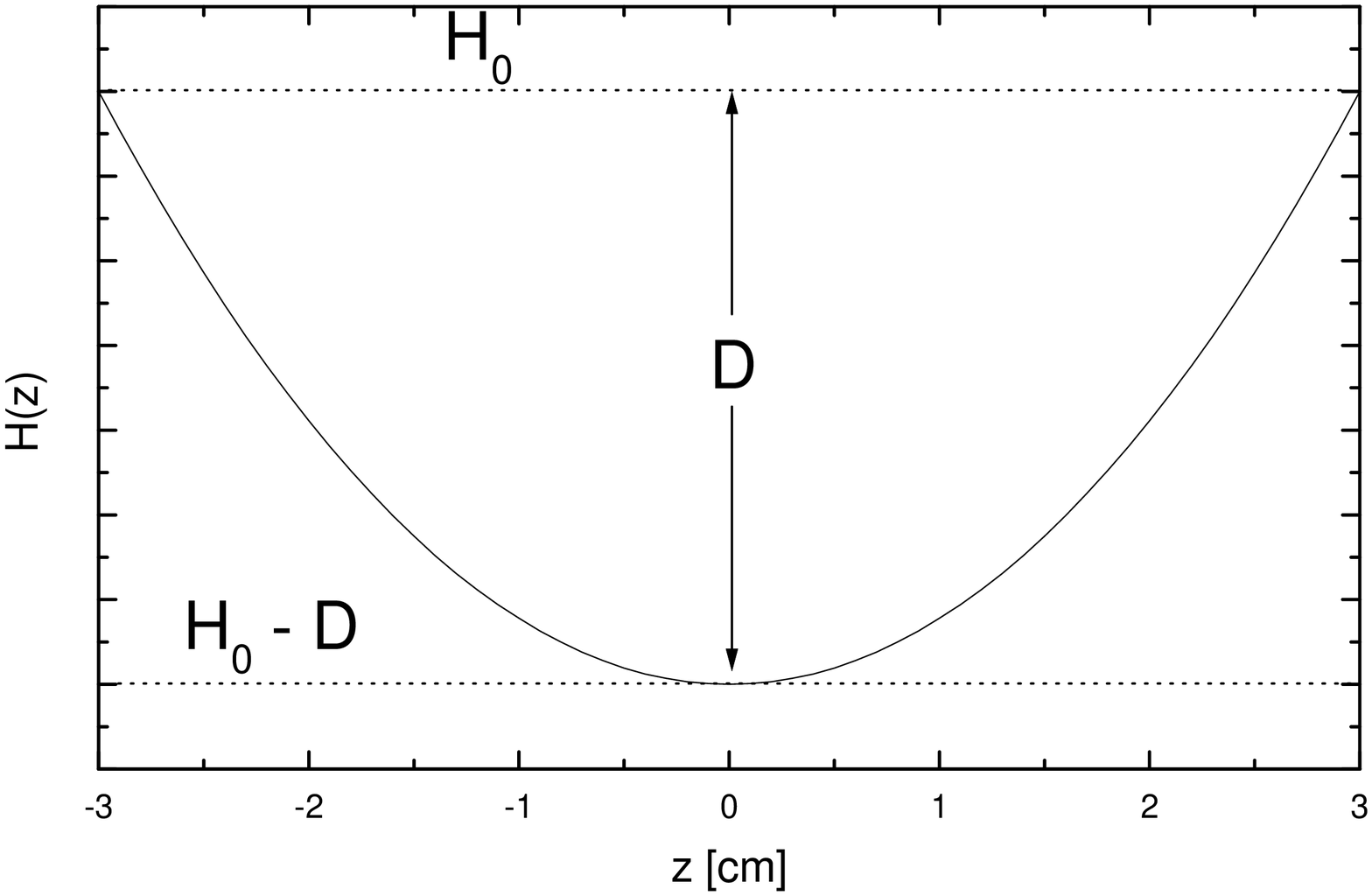}
\caption{The field profile assumed for the description of our data.}\label{fig:6}
\end{figure}

\begin{equation}
H(z)=D \left(\frac{z}{z_{0}}\right)^{2}+H_{0}-D \label{eq:4}
\end{equation}
where $H_{0}$ is the set value of the magnetic field and \textit{D} is the maximum field variation. The factor $z_{0}$ 
corresponds to the distance over which the field variation takes place, that is half the scan length (in our case 
3 cm).

\begin{figure}
\includegraphics[clip=true,width=75mm]{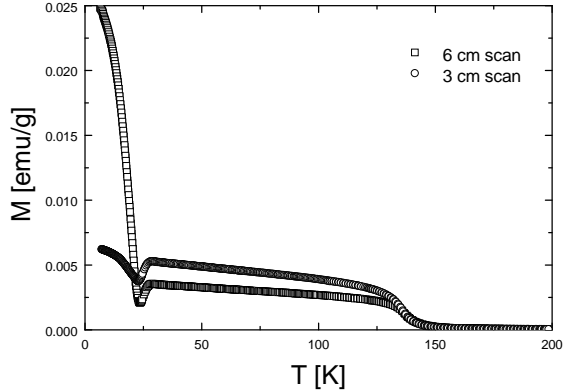}
\caption{f.c. d.c. magnetization measurement of Ru-1212 in a field of 50~$\mu$T after the cycle 0~T $\rightarrow$ 6~T $\rightarrow$ -6~T 
$\rightarrow$ 0~T $\rightarrow$ 50~$\mu$T using two different scan lengths.}\label{fig:7}
\end{figure}

We have used equations (16) and (17) from reference [11], following the procedure described in section 5 of that 
reference, in order to fit our measured \textit{V}(\textit{z}=0,\textit{T}) signal for the z.f.c. measurement of figure~\ref{fig:4} for 
$T<T_{c}$, assuming $T_{ir}=T_{c}=$30~K. The result is shown in the inset of figure~\ref{fig:4}. From the fit we obtain 
$n=2.32$ (see Eq.~\ref{eq:3}) and $r=D/H_{0}^{*}=0.145$.

From the value of the parameter \textit{r}, the change of the the field \textit{D} necessary to create artefacts in the d.c. 
magnetization measurements of Ru-1212 can be estimated, if the value of $H_{0}^{*}$ is known. It is known \cite{Poole}, 
that for a hysteresis loop, where the scanned field range is much bigger than $H^{*}$, like the ones at 25~K and 20~K in 
figure~\ref{fig:5}, $\mu_{0}M_{+}=\frac{1}{2}H^{*}$ 
or $\mu_{0}M_{-}=-\frac{1}{2}H^{*}$, where $M_{+}$ and $M_{-}$ correspond to the upper and lower magnetization plateau 
respectively. Thus, from the measured hysteresis curve at 25~K we find $H^{*}$=7.4~$\mu$T. Using Eq.~\ref{eq:3}, with 
the \textit{n} value as determined from the fitting of the \textit{V}(\textit{z}=0,\textit{T}) data, $H_{0}^{*}$ is determined as 
$H_{0}^{*}$=473~$\mu$T. 

For \textit{T}=20~K we 
calculate using Eq.~\ref{eq:3} $H^{*}$=37~$\mu$T, which is in very good agreement with the $H^{*}$=35~$\mu$T value determined 
from the measured hysteresis curve at 20~K. For \textit{T}=15~K though, Eq.~\ref{eq:3} gives $H^{*}$=90~$\mu$T. Thus, the scanned 
range of fields in figure~\ref{fig:5} is about two times the calculated value of $H^{*}$ at 15~K. This means \cite{Chen,Poole} that, 
if the power law used to calculate $H^{*}$ at 15~K was correct, the shape of the measured curve should be more flat, similar to those at 
25~K and 20~K. We have also used Eq.~\ref{eq:3} to estimate $H^{*}$ for \textit{T} = 10~K and 5~K and calculated $\mu_{0}M$ at characteristic 
points of the hysteresis loop using Table 12.2 in reference [12]. Deviations of the calculated values from the measured ones were observed in 
both cases. 
From this analysis we conclude that the power law of Eq.~\ref{eq:3} is valid only close to 
\textit{T$_{c}$} (15~K$<$\textit{T}$<$30~K). Indeed, deviations of the fitting curve from the measured \textit{V}(\textit{z}=0,\textit{T}) data 
were observed below 15~K.

With the calculated value of $H_{0}^{*}$ it is trivial to show, that the field change over the scanning length which caused the artefact shown in 
figure~\ref{fig:4} was $D=rH_{0}^{*}\sim$70~$\mu$T. Thus, field inhomogeneities less than 100~$\mu$T 
over the scaning length in the SQUID magnetometer are enough to create artefacts in the measurements of Ru-1212. This is already expected 
from figure~\ref{fig:5}, since the width of the hysteresis loops is less than 100~$\mu$T. We note, that field changes of the order of 100~$\mu$T, or 
even bigger, are not impossible for the remanent field of a superconducting magnet \cite{Blunt,McElfresh} and we expect, that measurements with set 
values of the magnetic field in the order of the remanent field, typical in the search for a Meissner state for Ru-1212, 
will not affect these changes. 

It is important to point out that the critical parameter which can create problems in the measurements of a superconducting sample 
is not the set value of the magnetic field but the absolute change of the field over the scan length. For a measurement in a set field of 
250~$\mu$T an inhomogeneity of 70~$\mu$T represents a field change of 28\% over the scan length. Nevertheless, this inhomogeneity will 
have the same effect in the measured properties of a superconducting sample even if the measurement is done in a set field of 1~T, where it 
represents a field change of only 0.007\%, which is much smaller than the field homogeneity claimed for the magnets 
of many of the commercially available magnetometers by their manufacturers. For Ru-1212 though, in high fields, a possible artefact in the signal related 
to the superconductivity of this compound will probably be hidden by the contribution from the Gd paramagnetic moments. 

It should not be assumed that artefacts in the d.c. magnetization measurements of Ru-1212 will always cause peak like features 
similar to those shown in figure~\ref{fig:4}. The specific characteristics of a possible artefact in a 
\textit{M}(\textit{T}) measurement will depend on the characteristics of the field profile in which the measurement was done. An
example for this is shown in figure~\ref{fig:7}. The first measurement of this Ru-1212 sample had shown a
peak like feature similar to those presented in figure~\ref{fig:4}. Nevertheless, after the magnet was cycled to high fields, 
a cycle very likely to have changed the field profile compared to that of the first measurement \cite{McElfresh}, a reversed peak was observed 
in the superconducting state of the sample, realised as a decrease of the magnetization just below \textit{T$_{c}$} followed by an increase 
of the magnetization at lower temperatures.
We could assume, that the reversed peak of figure~\ref{fig:7} is the result of a field profile, which is reversed 
compared to that of figure~\ref{fig:6}. This is in accordance with an observation made by McElfresh \textit{et al.} in reference [6], where 
(measured) symmetric field profiles with respect to a set value of the magnetic field created artefacts in the measured magnetic moment, which 
had the form of features reversed with respect to each other. 
Our assumption is additionally supported by the measurement with a smaller scan length of 3~cm, where we found that the measured magnetic 
moment in the normal state of the sample is now higher compared to the measurement with the 6~cm scan length. This indicates, that the average 
field for the measurement with the 3~cm scan length was higher than that for the measurement with the 6~cm scan length. Thus, the magnetic 
field for the measurements of figure~\ref{fig:7} can be assumed higher closer to the center of the magnet compared to its values close to the end points 
of the scan length. We note that many of the published d.c. magnetization data on Ru-1212 present 
features which seem to be reversed with respect to each other. For example the f.c. curves in reference [13] (we concentrate our attention 
to the field cooled curves because this is where someone would look for the Meissner effect) are similar to those of figure~\ref{fig:7} and 
present features reversed with respect to the curves presented in reference [14], which are similar to the curve in figure~\ref{fig:4}. 
Also the f.c. curve of Klamut \textit{et al.} \cite{Klamut}, with an increase of the magnetization just below \textit{T$_{c}$}, followed 
by a plateau at low temperatures is also reminiscent of a ``reversed'' effect compared to the f.c. curves published by
Bernhard \textit{et al.} \cite{Bernhard}, which show a decrease of the magnetization just below \textit{T$_{c}$} and a plateau at low temperatures.

In an attempt to 
eliminate the problems described above, small scan lengths are often used, in order to measure in the region of the magnet characterised 
by maximum uniformity. This tactic though, as shown in 
figure~\ref{fig:7}, does not always have the desired results. The smaller scan length in figure~\ref{fig:7} reduced the magnitude of the 
measured magnetic moment in the superconducting state of the sample but it did not eliminate the artefact. It should also be kept in mind, 
that small scan lengths extract a smaller proportion of the \textit{V}(\textit{z}) signal and can easier lead to mistakes since the centering 
of the sample relative to the gradiometer is now more critical. In principal, small scan lengths tend to degrade significantly the quality 
of the measured signal and in such cases an average of several scans is required to improve the statistics of the data.

\section{Concluding remarks} 

In summary, the quality of SQUID d.c. magnetization measurements on superconducting samples can drastically be affected if the measurements are 
done in an inhomogeneous field. 

It is possible that assymetric \textit{V}(\textit{z}) signals will warn the user of a SQUID magnetometer 
about possible problems. Nevertheless, in some cases the deformation of these signals is rather small and it will probably not be noticed. 

The sample's properties in combination with the level of field inhomogeneities will determine whether artefacts will appear in a d.c. magnetization 
measurement, the specific characteristics of which in a \textit{M}(\textit{T}) curve 
will depend on the specific field profile during the measurement. However, in some cases 
the level of field changes over the scan length required to create problems in a measurement is smaller than the claimed homogeneity for the 
magnets of many of the commercially available magnetometers.

The reduction of the scan length does not guarantee elimination of possible artefacts. It should also be kept in mind that the quality of the 
measured signal degrades significantly for short scan lengths. 

For the above reasons the validity of the data has to be carefully checked. The recording of the SQUID circuit's output voltage without moving the sample 
can give very useful information about what type of features are expected in the measurements of a superconducting sample. This is probably 
the most reliable test for the validity of the data since other tests, like checking the reproducibility of the data after the magnet is 
cycled to high magnetic fields (see figure~\ref{fig:7}), can lead to systematic errors if the magnetic field profile is reproduced. Even 
a controllable quench of the magnet, an option available with the most modern magnetometers, in order to eliminate effects related with the trapped flux 
in it can not guarantee high quality measurements. The increase of the current on the leads, after quenching, for the achievement of the desired 
field value, can cause, through the resulting field change, flux trapping effects, which will again affect the field homogeneity over 
the scanning length.     
Monitoring the SQUID circuit's output voltage independently of course requires that the magnetometer offers such an option. 

Testing the validity of the data is a crucial point since in some cases experimental artefacts can be described by impressive theories. For 
example, we have shown \cite{Papageorgiou} that the peak like features observed for Ru-1212 could be interpreted as an indication for the 
existence of the Paramagnetic Meissner Effect for this compound. By careful evaluation of the data though, we showed that this is not correct 
\cite{Papageorgiou}. Here we have shown that the Bean critical state model can be used for the description of the field penetration in Ru-1212 
in small magnetic fields. We have used this model to show that field changes less than 100~$\mu$T over the scanning length in the SQUID magnetometer 
are sufficient to create artefacts in the measured properties of this compound. Thus, careful evaluation of the data on this compound is 
always necessary. The observed sensitivity of the Ru-1212 measured magnetization in its superconducting state to the shape of the measuring field 
profile could serve as an explanation for the contradicting results on the magnetization of this compound in its superconducting state.

\end{document}